\documentclass[aps,prx,preprint,groupedaddress]{revtex4}\usepackage{tabulary,graphicx,amsmath,amsfonts,amssymb}
\usepackage[utf8x]{inputenc}
\usepackage{lineno} 
\usepackage{mathtools}
\usepackage{color}
\usepackage{subcaption}
\usepackage{cancel}
\definecolor{blueryb}{rgb}{0.01, 0.28, 1.0}

\definecolor{greenryb}{rgb}{0.1, 1.0, 0.1}

\usepackage{ragged2e}

\begin{document}

\title{Two state  model for the negative slope of the melting-curve}

\author{ Graeme J. Ackland$^{1*}$,  Hongxiang Zong$^{1,2}$, Victor Naden Robinson$^{1,3}$, 
Sandro Scandolo$^{3}$, and Andreas Hermann$^1$}

\affiliation{
$^1$ Centre for Science at Extreme Conditions and School of Physics and Astronomy, The University of Edinburgh, Edinburgh, EH9 3FD, UK
$^2$ State Key Laboratory for Mechanical Behavior of Materials, Xi’an  
Jiaotong University, Xi’an, Shanxi 710049, China
$^3$ The ``Abdus Salam'' International Centre for Theoretical Physics, I-34151 Trieste, Italy
$^*$ Corresponding authors: gjackland@ed.ac.uk
}

\begin{abstract}
    We present a thermodynamic model which explains the presence of a negative slope in the melt curve, as observed in systems as diverse as the alkali metals and molecular hydrogen at high pressure. We assume that components of the system can be in one of two well defined states - one associated with low energy, the other with low volume.
    The model exhibits a number of measurable features which are also observed in these systems and are therefore expected to be associated with all negative Clapeyron-slope systems: first order phase transitions, thermodynamic anomalies along Widom lines.
    The melt curve maximum is a feature of the model, but appears well below the pressures where the change in state occurs in the solid: the solid-solid transition is related to the melt line minimum.  An example of the model fitted to the electride transition in potassium is discussed.
\end{abstract}

\maketitle 


\section{Introduction}

Improvements in high pressure and temperature experiments mean that the topic of liquid-liquid phase transitions has attracted significant attention recently.   In particular, there are debates about whether a change in liquid structure can be regarded as a true phase transition, or a gradual crossover.

Determining this is challenging for either experiment or simulation.  In a diamond anvil cell experiment it is near-impossible to observe phase coexistence and be confident that the system has reached thermodynamic equilibrium.  Indeed, many if not most high pressure experiments report phase coexistence across a range of pressures - something which is forbidden in equilibrium thermodynamics.   The situation is no different in simulations - typical electronic structure calculations are done at a given pressure and temperature and ``discontinuities" are inferred by extrapolation or, at best, hysteresis.

The melt curve for most materials has a positive slope on a PT phase diagram. This means that the liquid is less dense than the solid.
There are exceptions, notably, water is denser than ice, and other examples among elements include silicon, gallium, carbon.
These textbook exceptions at ambient pressure can generally be understood as due to the partial breakdown of a network of well defined covalent or hydrogen bonds causing the atoms in the liquid to have a higher coordination than the solid.

Another group of materials which have a negative Clapeyron slope at high pressure are the alkali metals\cite{Tsuji1990,Tsuji1996,falconi2006X-ray,mcbride2015one,marques2009potassium,gatti2010sodium,woolman2018structural,Marques2011,marques2011optical,zhao2019commensurate,Miao2014,frost2018equation,frost2019high,schaeffer2012high,guillaume2011cold,gregoryanz2008structural,eshet2012microscopic,hong2019reentrant,zha1985melting,narygina2011melting,kartoon2021structural}.  Here, the slope is typically positive at ambient pressure, reaching a maximum, then becoming negative in a pressure region where the solid phase is typically close packed.  At still higher pressures, there is a minimum in the melt curve before the slope becomes positive again.
The solid phase in the region of negative slope is close packed, so the densification on melting cannot come from a collapsing open network.   
Curiously, hydrogen has a remarkably similar phase diagram to the alkalis which can be explained by competition between free rotors and quadrupole interactions\cite{magduau2017simple,zong2020understanding,van2020quadrupole}

Density functional theory can reproduce the negative slope\cite{hernandez2010first,feng2015nuclear,robinson2019chain}. It also shows some anomalous behaviour in liquid heat capacity, compressibility, viscosity, and thermal expansion\cite{zongK}. This implies that there is some significant change in the liquid binding, though whether it is a true transformation or a crossover remains unclear. 
As a consequence, there is renewed interest is analytic equation of state which can be fitted to data.  For single phases, functional forms such as the Vinet equation of state work well, but  many interesting phenomena occur where the equation of state is concave or discontinuous.

The thermodynamically stable state is the one with the lowest Gibbs free energy:

\[  G(P,T) = U + PV - TS \]

taking $P$ and $T$ as the independent variables. Evidently $G$ depends on three quantities, energy, density and entropy.  Any attempt to relate microscopic to macroscopic properties needs to consider all three, and how they vary between phases.

There are a wide range of approaches to describe complex high pressure structures.  Those based on electronic-structure include electride\cite{marques2009potassium}, two-band\cite{ackland2003two,ackland2006two}, Fermi-surface\cite{Degtyareva2003,ackland2004origin,degtyareva2006simple},  s-p\cite{jr:Yanming} or s-d\cite{ReedBa2000} transfer, Mott transitions\cite{johansson1974alpha} or pairing\cite{jr:Pairing}, molecularisation\cite{ashcroft2000hydrogen,katayama2000first,monaco2003nature}, polymerisation, \cite{ross2006polymerization}  and ``simple-complex"\cite{gorelli2018simple} transition types.
Other approaches based on interatomic forces include soft-core\cite{ogura1977computer,young1977melting}, and  associating particles\cite{Jackson_etal_1988,Hopkins_etal_2006}.
 
Despite this huge variety of microscopic models, we are unaware of a simple, analytic thermodynamic model for the melting point maxima\cite{rapoport1968melting,gregoryanz2005melting,kechin2001melting} and liquid-liquid transformation\cite{mukherjee2007high, Cadien2013}.  Rapoport\cite{Rapoport1967} implies that Klement built such a model, but it was never published - Rapoport's own analysis of Klement's model does not show a melting point maximum.  A number of lattice-based approaches have been tried \cite{dijkstra1994evidence}, but for obvious reasons their applicability to the fluid state is debatable. 

We note that most of the microscopic models are based on a trade-off between two types of interaction, one which has lower energy, the other lower volume.
The purpose of this paper is to lay out the minimal requirements for an analytic model of a discontinuous liquid-liquid transformation and a melting point maximum based only on only the idea that a material can adopt two different  states. 

The paper is structured as follows - we start by deriving thermodynamic results for heat capacity, expansivity and compressibility in a convenient analytic form.  We then present a mixing model between two thermodynamic states, demonstrating the Widom lines.  A microscopic model inspired by the electride transition\cite{dye1990electrides,ahulwalia1999structural,marques2009potassium,pickard2009dense,gatti2010sodium,raty2007electronic,rousseau2011exotic,li2015metallic,woolman2018structural,Marques2011,marques2011optical,zhao2019commensurate,Miao2014,miao2015high,yu2018optical,frost2018equation,elatresh2019high,paul2020thermal,ayrinhac2020high}, where the states differ only in volumes, is worked through in detail for both solid and liquid cases.
It is shown that this model is sufficient to obtain the melt curve maximum, and can support a discontinuous phase transition in the liquid.
A parameterization for potassium is presented.

\section{Thermodynamic model}
The theory derived here is of very general applicability.  However, we found it helpful to have a concrete microscopic model in mind as it is developed. 

\subsection{Motivation from simple metals at pressure}

We propose that the structure of the high-pressure alkali metals can be modeled as a mixture of two distinct electronic states: a low-pressure  s-type free electrons, and a high-pressure ``electride’’ state, with electrons localized in interstitial pockets, referred to as pseudoanions.  In the case of fcc, we can imagine that the octahedral site is the pseudoanion, so the electride has a rocksalt structure. 
This should not be taken too literally because in reality, the electride transition is accompanied by a crystal structural transformation.  Similar evolution happens in a liquid, but here the transition is continuous because  differently-sized pseudoanion sites are available, and there is no symmetry.
This microscopic picture can be related to a macroscopic one by considering the energy, volume and entropy of the two states:

\begin{itemize}
    \item The electride has higher energy because the electron is confined away from the positively charged ion. 
    \item The electride has small volume, because it can occupy the interstitial site between ions, leading to more efficient packing.
    \item A mixture of the two states gives higher entropy. 
\end{itemize}

In addition to the electride transition, we may also compare solid and liquid phases for which
the solid has lower entropy and enthalpy, independent of the electride fraction.

The need to describe $U$, $TS$ and $PV$ for each phase means that even the simplest model inevitably has several parameters.

\subsection{Thermodynamics}
In a general two-state model, a Gibbs free energy is written as $G(x,P,T)$ where $x$ is the fraction of one of the two states.  The equilibrium value for $G(P,T)$ is obtained by minimising $G(x,P,T)$ with respect to $x$. So for all $P,T$ we have

\begin{equation} G(P,T)= \min_x G(x,P,T) \label{eq:Gx}
\end{equation}
A necessary, but not sufficient requirement for equilibrium is: 
\begin{equation} \left (\frac{\partial G}{\partial x}\right )_{P,T}=\left (\frac{\partial H}{\partial x}\right )_{P,T} -T \left (\frac{\partial S}{\partial x}\right )_{P,T} =0 \label{eq:eqm} \end{equation}
Simply solving that equation will also generate unphysical free energy maxima, and metastable states.

Thermodynamic properties are obtained as derivatives of the free energy. Although the calculus is routine, we present the results here because of the additional terms which arise due to the $x$ factor, and the fact that some derivative cannot be written analytically because of the requirement to minimise $x$.

\subsubsection{Heat capacity $C_p$}

The standard thermodynamic definitions of the heat capacity are 

\begin{equation} C_p = \left (\frac{\partial H}{\partial T}\right )_P
= T\left (\frac{\partial S}{\partial T}\right )_P=T\left
(\frac{\partial^2 }{\partial T^2}Min_{x} [G(P,T,x)]\right )_P
\end{equation}

Note that $x$ is not an independent variable, and changes in $x$ contribute to the heat capacity.
\[ C_P =\left ( \frac{\partial H}{\partial T}\right )_{P,x} + 
\left ( \frac{\partial H}{\partial x}\right )_P\left ( \frac{\partial x}{\partial T}\right )_P\]  

The quantity $\frac{\partial T}{\partial x}$ can be awkward to evaluate, so to eliminate it, we consider

\begin{equation}\left (\frac{\partial}{\partial T}\right )\left (\frac{\partial G}{\partial x}\right ) = -\left (\frac{\partial S}{\partial x}\right ) + \left (\frac{\partial x}{\partial T}\right ) \left[ \left (\frac{\partial^2 G}{\partial x^2}\right )
\right]
\end{equation}

dropping the subscripts for clarity.  Using the equilibrium condition (Eq.\ref{eq:eqm}), this gives

\begin{equation} \left ( \frac{\partial T}{\partial x}\right )_P
= T \left (\frac{\partial^2 G}{\partial x^2}\right )_P / \left (\frac{\partial H}{\partial x}\right )_P \label{eq:Tx}
\end{equation}

From which the expression for the heat capacity becomes:
\begin{equation} C_P =\left ( \frac{\partial H}{\partial T}\right )_{P} 
= \left ( \frac{\partial H}{\partial T}\right )_{P,x} + \frac{1}{T}\left (\frac{\partial H}{\partial x}\right )^2_P /  \left (\frac{\partial^2 G}{\partial x^2}\right )_P 
\end{equation}

\subsubsection{Isothermal Compressibility}

The standard thermodynamic definitions of compressibility are 

\begin{equation} K = -\frac{1}{V}\left ( \frac{\partial V}{\partial P}\right )_T
= -\frac{1}{V}\left (\frac{\partial^2 G}{\partial P^2}\right )_T
\end{equation}

including the internal variable $x$.
\[ \left (\frac{\partial^2 G}{\partial P^2}\right )_T = 
\left (\frac{\partial^2 G}{\partial P^2}\right )_{T,x} +
2 
\left (\frac{\partial^2 G}{\partial P\partial x}\right )_T
\left (\frac{\partial x}{\partial P}\right )_{T} + \cancelto{0}{
\left (\frac{\partial G}{\partial x} \right )_T\left (\frac{\partial^2 x}{\partial P^2}\right )_T}
\]
The equilibrium condition $\left ( \frac{dG}{dx}\right ) =0$ ensures that the third term is zero.

Again, there is no convenient relationship between $P$ and $x$, but following a similar argument to Eq.\ref{eq:Tx} we find

\[\left ( \frac{\partial x}{\partial P}\right )_T
=-\left (\frac{\partial^2 G}{\partial P\partial x}\right )_T/\left (\frac{\partial^2 G}{\partial x^2}\right )_T
\]
and

\[\left ( \frac{\partial^2 x}{\partial P^2}\right )_T
=\left (\frac{\partial^3 G}{\partial P^2\partial x}\right )_T/\left (\frac{\partial^2 G}{\partial x^2}\right )_T
-\left (\frac{\partial^3 G}{\partial P\partial x^2}\right )^2_T/\left (\frac{\partial^2 G}{\partial x^2}\right )^2_T
\]

\subsubsection{Thermal Expansion}

The standard thermodynamic definitions of compressibility are 

\begin{equation} \alpha = \frac{1}{V}\left ( \frac{\partial V}{\partial T}\right )_T
= \frac{1}{V}\left (\frac{\partial^2 G}{\partial T \partial P}\right )
\end{equation}
 
 \begin{equation} 
 \frac{\partial^2 G}{\partial T \partial P} = \left (\frac{\partial^2 G}{\partial T \partial P}\right )_x + 
 \left (\frac{\partial^2 G}{\partial P\partial x}\right )_T
\left (\frac{\partial x}{\partial T}\right )_{P} 
+\left (\frac{\partial^2 G}{\partial T\partial x}\right )_P
\left (\frac{\partial x}{\partial P}\right )_{T} 
\end{equation}

Again, using the equilibrium condition
$\left (\frac{\partial G}{\partial x} \right )_T=0$.

\subsection{Linear combination model with ideal solution\label{model:SS}}

In a slightly more specific model, a system is described by particles in two possible thermodynamic states A (x=1) and B (x=0). When mixed in an ideal solution, the Gibbs free energy is given by:
\begin{equation}
    G(P,T) = x G(1,P,T) + (1-x) G(0,P,T) + RT[x \ln{x} + (1-x)\ln{(1-x)}] 
    \label{Gibbs:SS}
\end{equation}

where $x$ is the fraction of particles in state A, $G_A=G(1,P,T)$ and $G_B=G(0,P,T)$ are the Gibbs free energies of pure A and B states. 
The equilibrium value for $x$ is found by minimising  $G(P,T)$:
\begin{equation}
    x(P,T) = \frac{e^{-G_A/RT}}{e^{-G_A/RT}+e^{-G_B/RT}} 
    = \frac{1}{1+e^{-\Delta G/RT}} 
    \\
    \label{xp:SS}
\end{equation}
with $\Delta G= G_B-G_A$.

We can find the volume
\begin{equation}
    V(P,T) = \left ( \frac{\partial G}{\partial P}\right )_T = xV_A(P,T) + (1-x)V_B(P,T) 
    \label{volume:SS}
\end{equation}
and entropy
\begin{equation}
    S(P,T) = -\left ( \frac{\partial G}{\partial T}\right )_P = xS_A(P,T) + (1-x)S_B(P,T) +   R[x \log{x} + (1-x)\log{(1-x)}] 
    \label{entropy:SS}
\end{equation}
always remembering that $x=x(P,T)$.
We further derive analytic expressions for compressibility
\begin{equation}
    \kappa_T = -\frac{1}{V}\left (\frac{\partial V}{\partial P}\right )_T=  x \kappa_{T,A} + (1-x)\kappa_{T,B} +
        \frac{\Delta V}{V} \left ( \frac{\partial x}{\partial P}\right )_T
    \label{kappa:SS}
\end{equation}
with $\Delta V$=$V_B-V_A$
and thermal expansion
\begin{equation} 
    \alpha = \frac{1}{V}\left( \frac{\partial V}{\partial T}\right )_P =  x \alpha_{T,A} + (1-x)\alpha_{T,B} +
        \frac{\Delta V}{V} \left ( \frac{\partial x}{\partial T}\right )_P
    \label{alpha:SS}
\end{equation}
Both of which have an anomalous component arising from the conversion of material between the two states, in addition to the weighted average.
For the heat capacity there is an additional anomalous term from the mixing entropy
\begin{equation}
    C_P = T \left ( \frac{\partial S}{\partial T}\right )_P = x C_{P,A} + (1-x)C_{P,B} +\left (
\Delta S      +  R\ln{\frac{x}{1-x}} \right ) \left ( \frac{\partial x}{\partial T}\right)_P 
    \label{cp:SS}
\end{equation}
with $\Delta S$=$S_B-S_A$.

 From Equation \ref{xp:SS} we immediately see that there is no discontinuity in $x$, from which it follows that this model cannot describe a phase transition, only a crossover.  We also observe that the ideal solution entropy ensures that mathematically, as well as intuitively, $0<x<1$.

\subsection{Non-ideal solution solid model}

A small rephrasing of the Bragg-Williams\cite{bragg1934effect} (BW) model can be used to extend the model from section \ref{model:SS} to describe a discontinuous transition within a single solid phase.  BW is a mean field approximation to the Ising model, where for high-pressure applications the spins are mapped to ``electride" and ``s-electron" states, and the ``field" is mapped to the enthalpy difference between the two states. Although the model has wider applicability, will use the electride terminology here.

The enthalpy in this case is 
\[ H=x(\Delta U_e + P\Delta V_e) +Jx(1-x)\]

where $x$ is the electride fraction,  $\Delta U_e$ and  $\Delta V_e$ are the change in energy and volume when an electron moves to an electride pseudoanion site, both assumed positive,  $J$ is a local coupling between electride and free electron. 
A high pressure phase transition at $T=0$ occurs when the field/enthalpy difference changes sign ($P_T=\Delta U/\Delta V$).

Including entropy, the Gibbs Free Energy is:
\begin{equation}
G_{BW} = x(\Delta U_e - P \Delta V_e -T \Delta S_e) + RT[x\ln{x}+(1-x)\ln(1-x)] +Jx(1-x) 
\label{gibbs:BW}
\end{equation}

We find that the $x$-dependent contributions are
\[ V = -x\Delta V_e \]
\[ U = x \Delta U_e + Jx(1-x) \]
\[ S =  R[x\ln{x}+(1-x)\ln(1-x)] +\Delta S_e
\]
Obtaining these results by differentiating $G$ is not completely trivial, as they rely on 
the stationary property of $G(x)$ at equilibrium (Eq.\ref{eq:eqm}).

We now find

\begin{equation}
\frac{\partial H}{\partial x} = (\Delta U_e - P \Delta V_e)  +J(1-2x) 
\end{equation}
\begin{equation}
\frac{\partial G}{\partial x} = (\Delta U_e - P \Delta V_e - T\Delta S_e) + RT\ln[{x/(1-x)}] +J(1-2x) 
\end{equation}
\begin{equation}
\frac{\partial^2 G}{\partial x^2} =  \frac{RT}{x(1-x)} -2J 
\end{equation}

\begin{equation} C_P(x) = \frac{1}{T}
\frac{\left (
 \Delta U_e - P \Delta V_e  +J(1-2x)\right )^2 
}{ \frac{RT}{x(1-x)} -2J  }
\end{equation}

We can see immediately that the heat capacity has a discontinuity if $RT/2J= x(1-x)$,
and since $x(1-x)$  must lie between 0 and 1/4, a discontinuous phase transition occurs for any $T<J/2R$ 
at $P=\Delta U_e/\Delta V_e$. Interestingly, along a line above the critical point, the contribution to $C_P$ goes to zero

\begin{figure}[!htb]
\centering
\includegraphics[width=0.98\columnwidth]{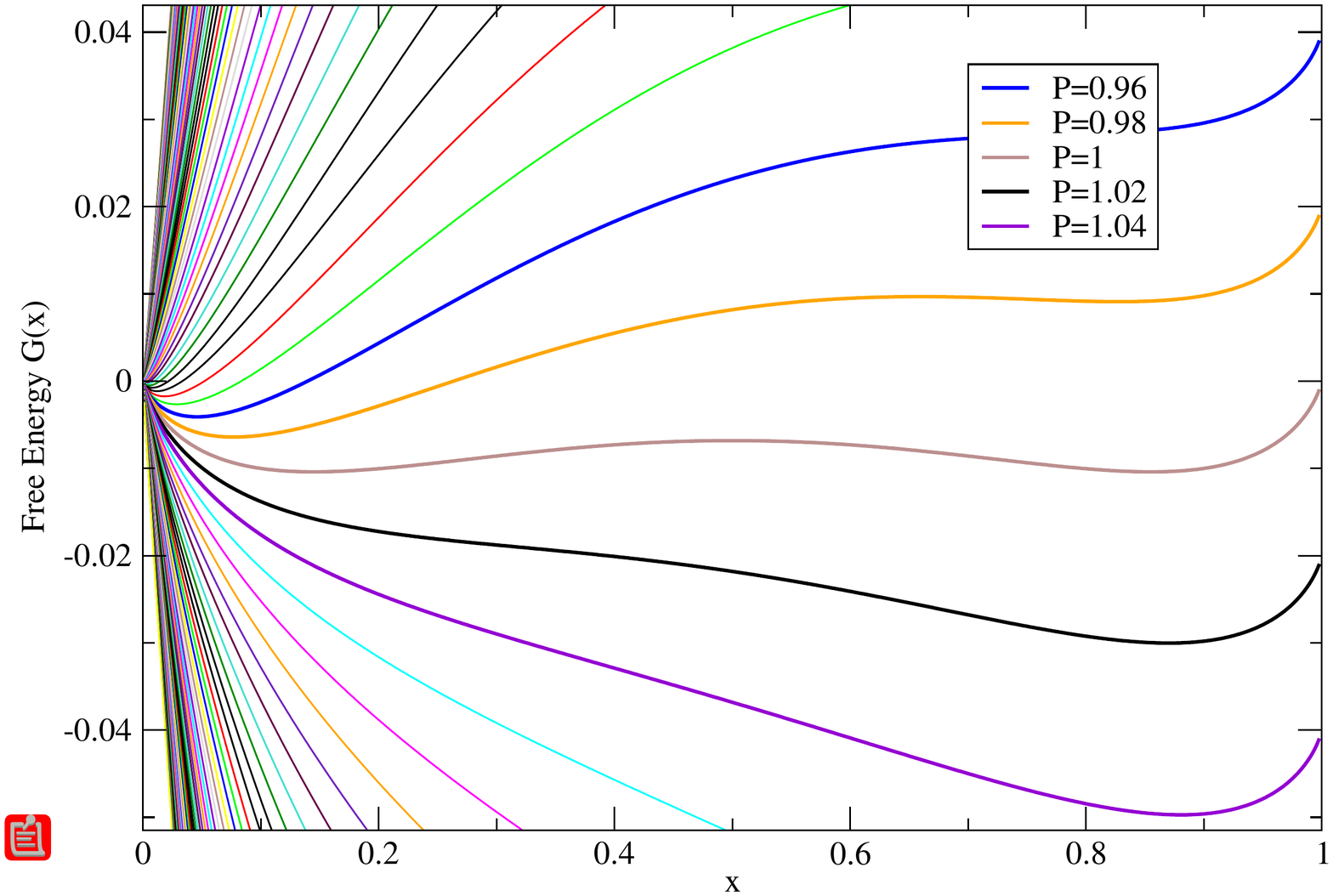}
\includegraphics[width=0.98\columnwidth]{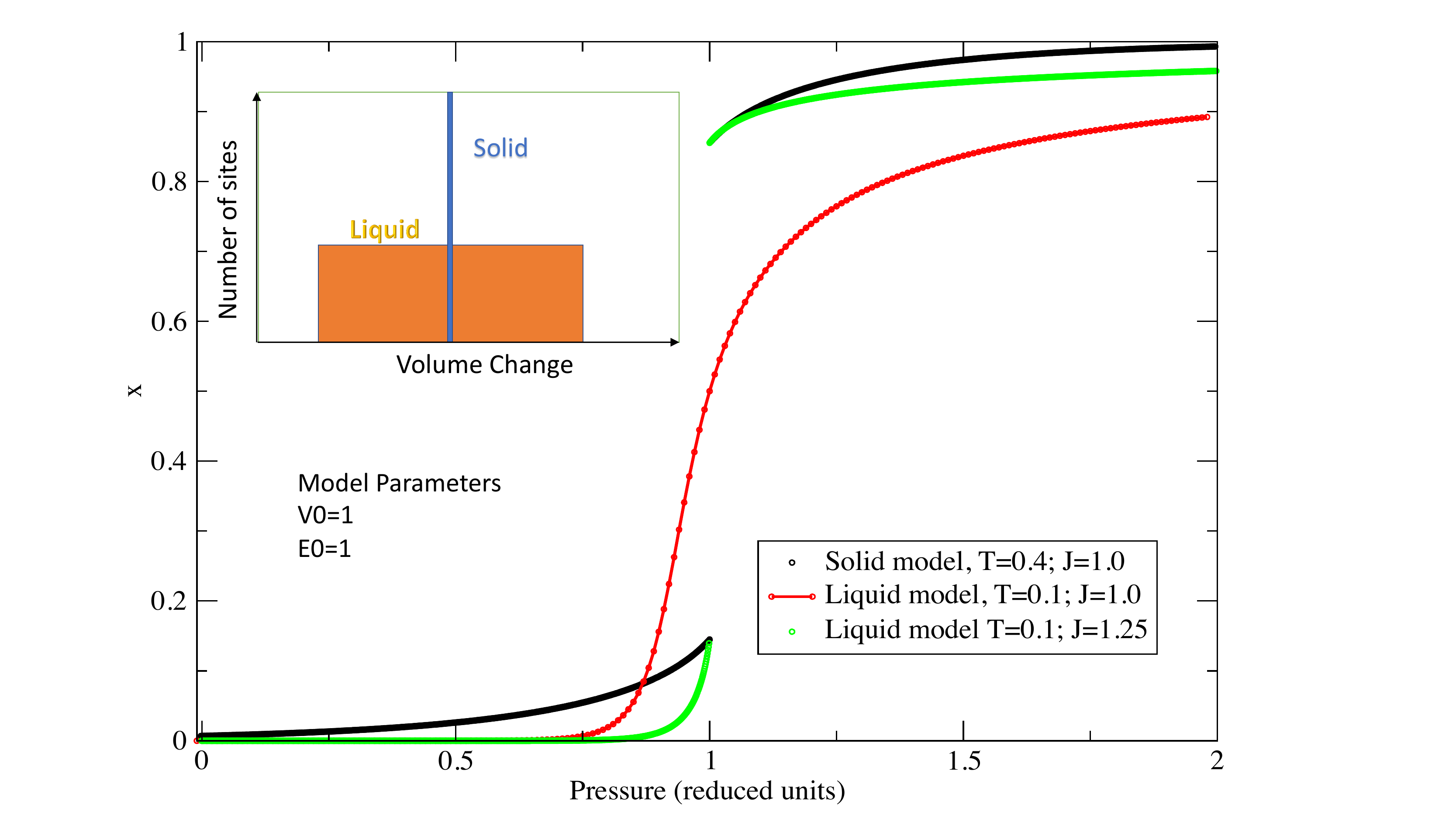}
\caption{(top) Variation of free energy with x for liquid model with $\Delta V_0=1$, $\Delta E_0=1$, J=1.25 at T=0.1. Lines correspond to different pressures with selected values around the transition at P=1 highlighted.  The equilibrium value of $x$ corresponds to the free energy minimum.  These conditions permit a first order transformation, and a metastable state can be seen for P=0.98. 
(bottom) Variation of fraction of two states ($x$) for solid and liquid model, with identical energy and volume differences.  At reduced T=0.4 the solid already exhibits a discontinuous phase transition, while the liquid does not (T=0.1 shown).  If J is increased from 1.0 to 1.25, the liquid model also exhibits a phase transition. 
The chosen values of $\Delta E_0=1$; $\Delta V_0=1$ mean that the transition pressure is at $P=1$ in either model.
Inset - schematic showing the different volume changes available in liquid (Orange) compared with unique value in solid (blue, delta function).\label{fig:BWx}}
\end{figure}

If it seems odd that $V$ and $U$ go to zero, at high and low pressures. remember that the
full free energy of the system will include terms independent of $x$, representing the equation of state of a reference $(x=0)$ material. 
  To compare with a real system, one needs to add an $x$-independent free energy $G_{ref}(P,T)$ to Eq.~\ref{gibbs:BW} which adds a smoothly varying additional term to all quantities.

Bragg and Williams considered an atomic level system\cite{bragg1934effect}, so assumed that the two sites have equal entropy; In applications such as polymerization or atomic-molecular where the number of independent objects changes a further term $T\Delta S_e=0$ could be added to relax this assumption.  This introduces a slope to the phase boundary and additional tilt to the Widom line, but does not change the general picture.

\subsection{Two site electride liquid model}

Our liquid more differs from the solid in just one detail.
Because of the symmetry in a crystal, the available electride sites are equivalent. 
In a liquid, we assume there are a range of different possible electride sites, each entailing different volume changes $\Delta V$ (see fig.\ref{fig:BWx}).

The site offering the largest volume reduction will be  occupied first. For simplicity, we assume sites are linearly distributed and range from $-2 \Delta V_e $ to 0.  Note that a positive $\Delta V$ implies that the electride would increase the volume, so such sites will never be occupied.

With finite electride fraction $x$, the total volume therefore changes by:

\begin{equation} \Delta V = \int_0^x 2\Delta V_e (x' - 1) dx' = \Delta V_e (x^2-2x)
\end{equation}

The excess Gibbs Free Energy is thus:
\begin{equation}
G_{EL}(x) = x\Delta U_e - x(2-x)P \Delta V_e + k_B T[x\ln{x}+(1-x)\ln(1-x)] +Jx(1-x) 
\end{equation}

There are now nonlinearities in  energy, entropy and density. 
\[ V = -x(2-x)\Delta V_e \]
\[ U = x \Delta U + Jx(1-x) \]
\[ S =  R[x\ln{x}+(1-x)\ln(1-x)]
\]

We now find

\begin{equation}
\frac{\partial H}{\partial x} = (\Delta U_e - (2-2x)P \Delta V_e)  +J(1-2x) 
\end{equation}
\begin{equation}
\frac{\partial G}{\partial x} = (\Delta U_e - (2-2x)P \Delta V_e) + RT\ln[{x/(1-x)}] +J(1-2x) 
\end{equation}
\begin{equation}
\frac{\partial^2 G}{\partial x^2} = 2P\Delta V_e + \frac{RT}{x(1-x)} -2J 
\end{equation}

\begin{equation} C_P = \frac{1}{T}
\frac{\left [
 \Delta U_e - (2-2x) P \Delta V_e  +J(1-2x)\right ]^2 
}{ \frac{RT}{x(1-x)} -2J +2P\Delta V_e  }
\end{equation}

This model does not necessarily have a critical point: the entropic and volume terms are always convex, so only the demixing $J$ term can drive phase separation.  Whether this happens depends on the value of $P$ at the putative phase boundary $x=\frac{1}{2}$, giving $P=\Delta U_e/\Delta V_e$.

\begin{figure}[ht]
\centering
\includegraphics[width=0.98\columnwidth]{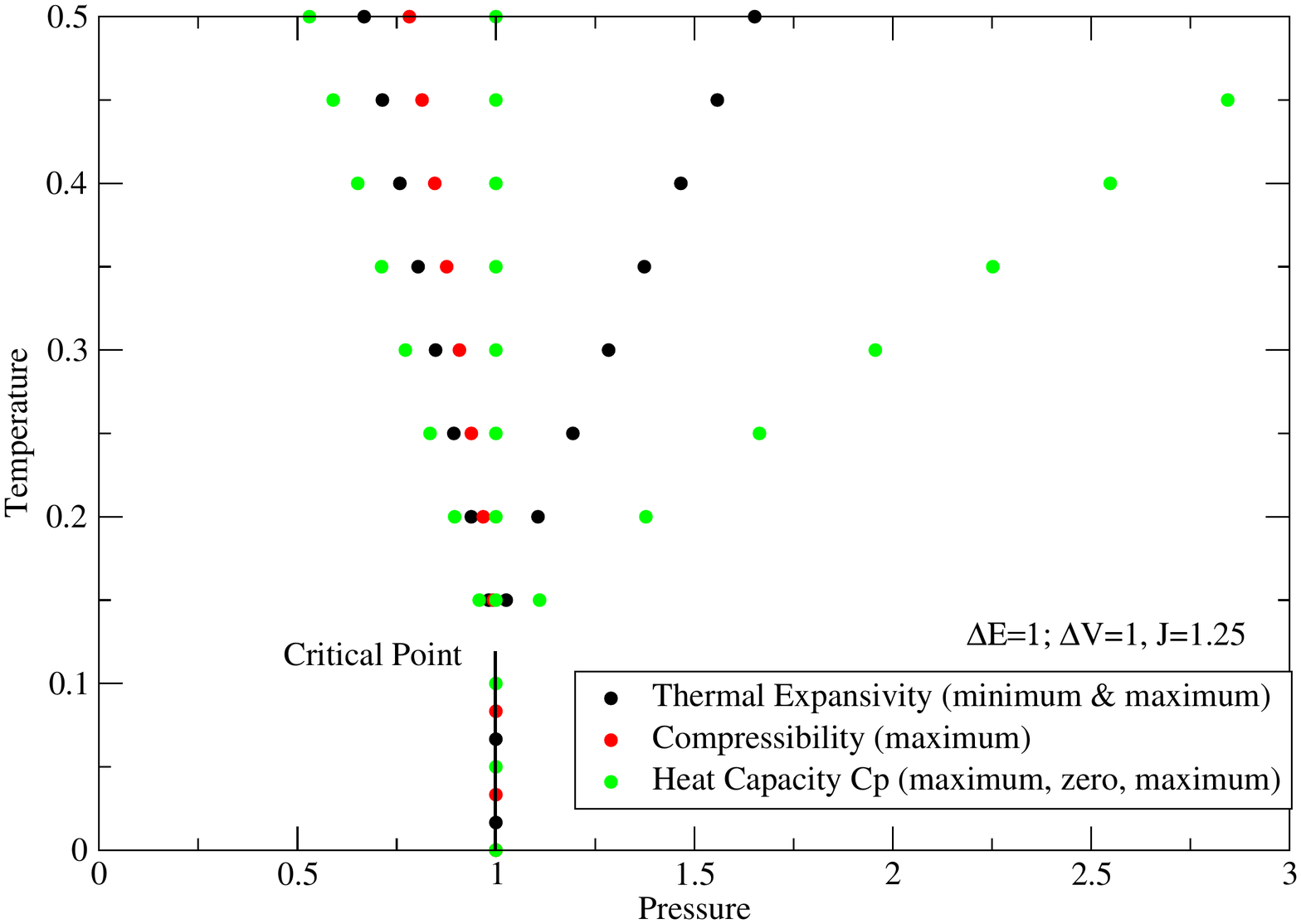}
\caption{Phase diagram for the liquid model with parameters as shown ($\Delta S=0$ - if non-zero then the phase transition line has a slope.  Points are the calculated maxima and minimum of the thermodynamic anomaly in compressibility, expansivity and heat capacity. The first order transition between high and low $x$ liquids ends in a critical point: in application of the model to real materials, this critical point may lie below the melt line.  Above the critical temperature, the extrema of the thermodynamic properties trace out the Widom lines which converge and end at the critical point.}
\label{fig:Widom}
\end{figure}

\begin{figure}[ht]
\centering
\includegraphics[width=0.48\columnwidth]{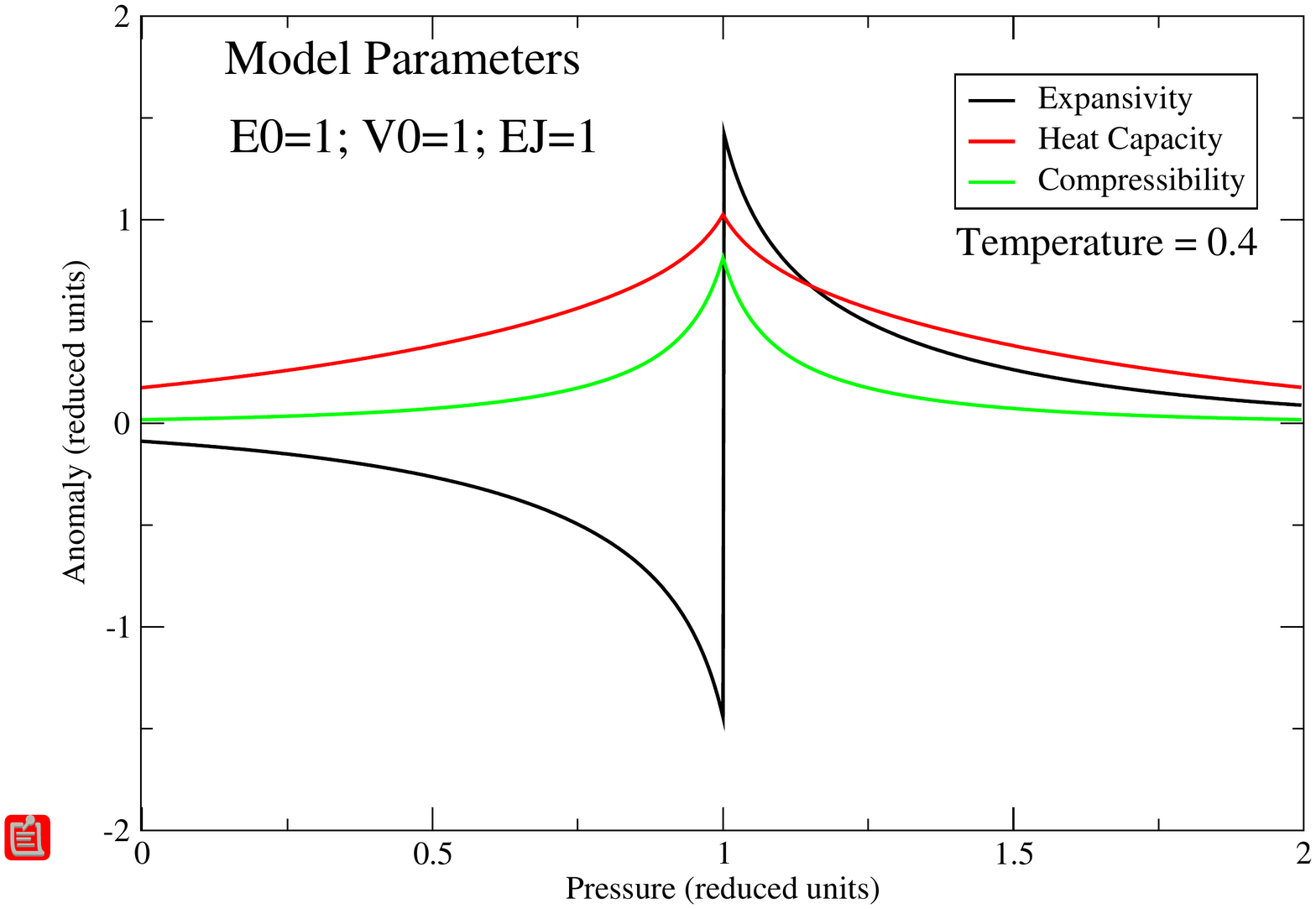}
\includegraphics[width=0.48\columnwidth]{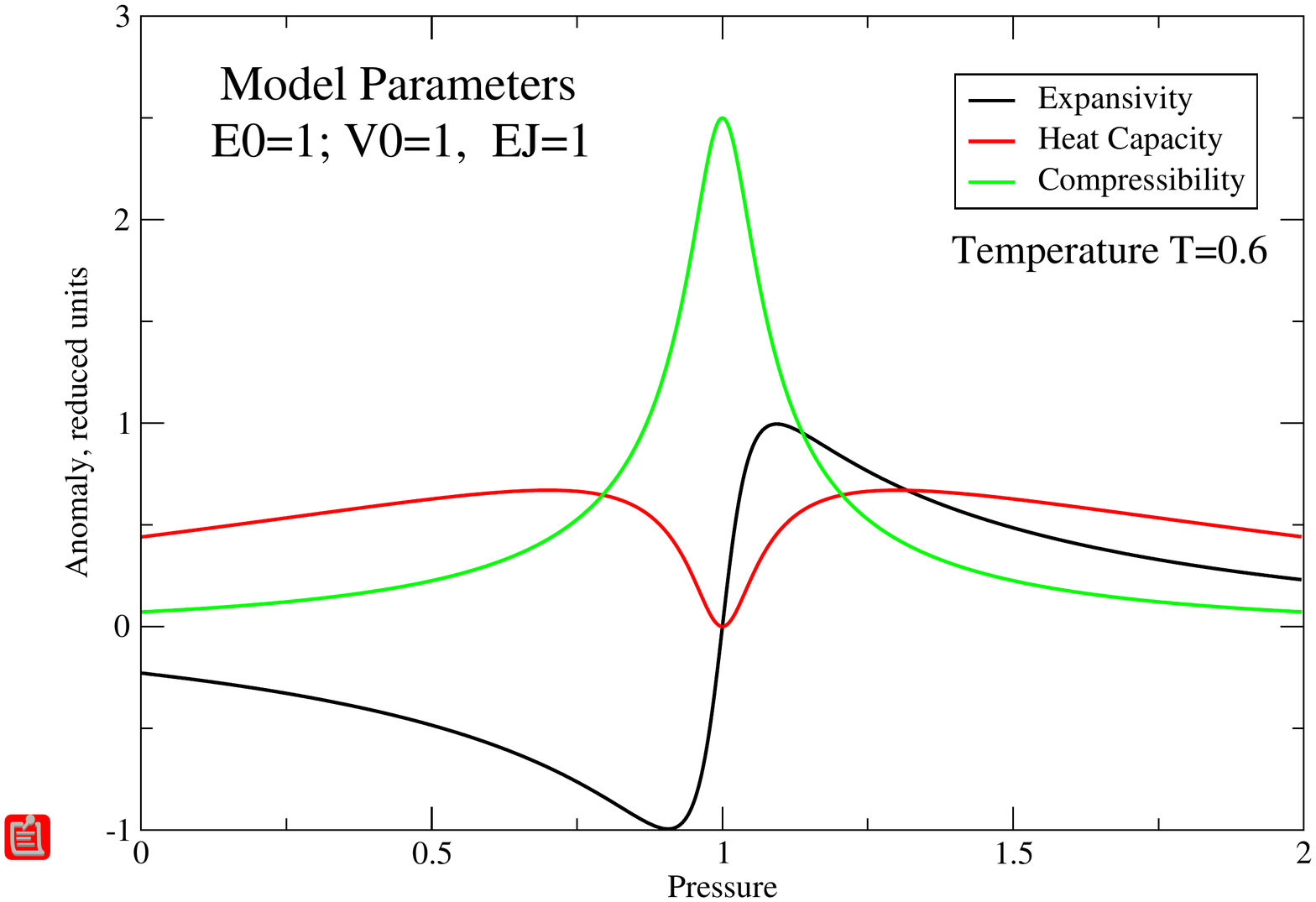}
\includegraphics[width=0.48\columnwidth]{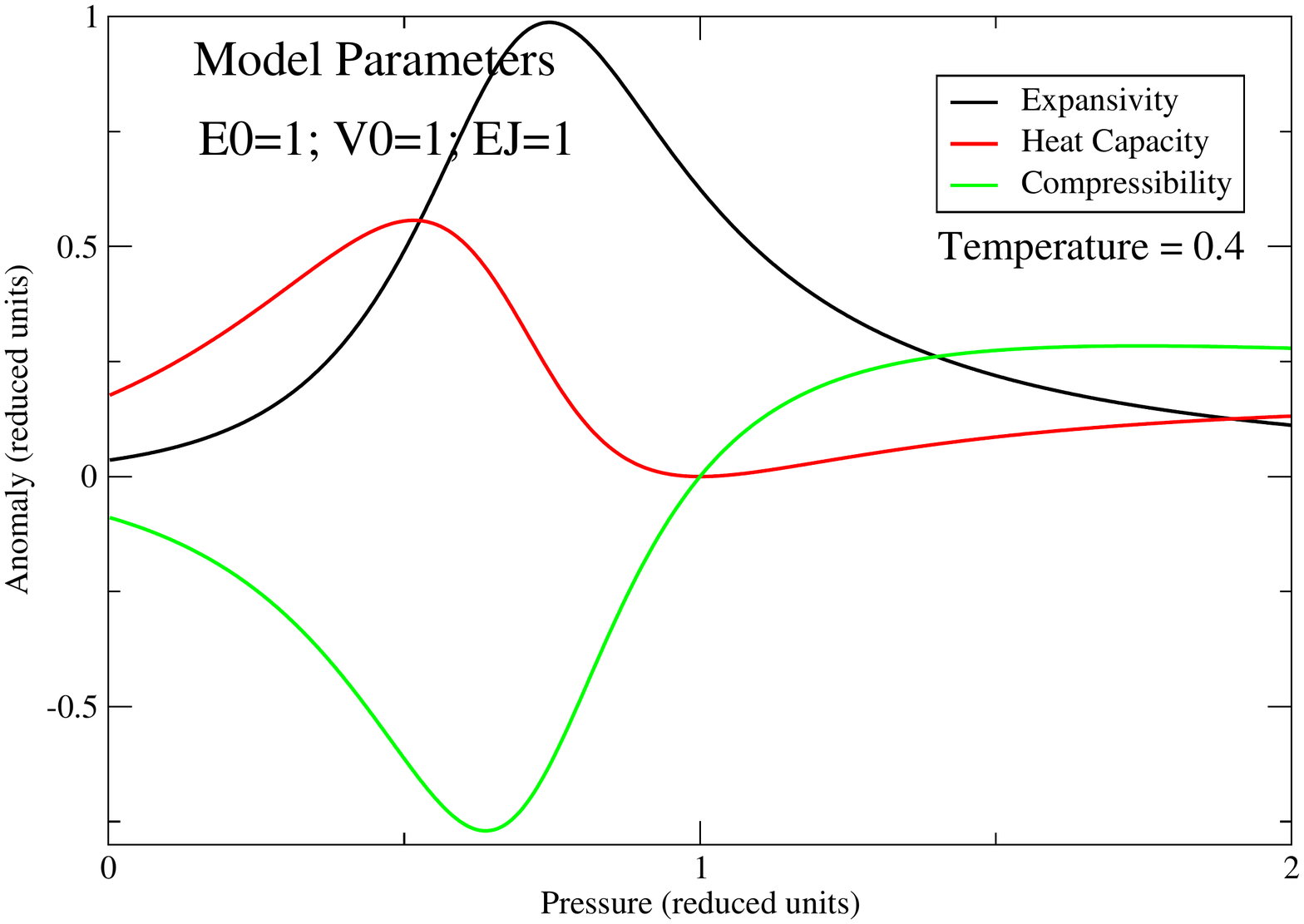}
\includegraphics[width=0.48\columnwidth]{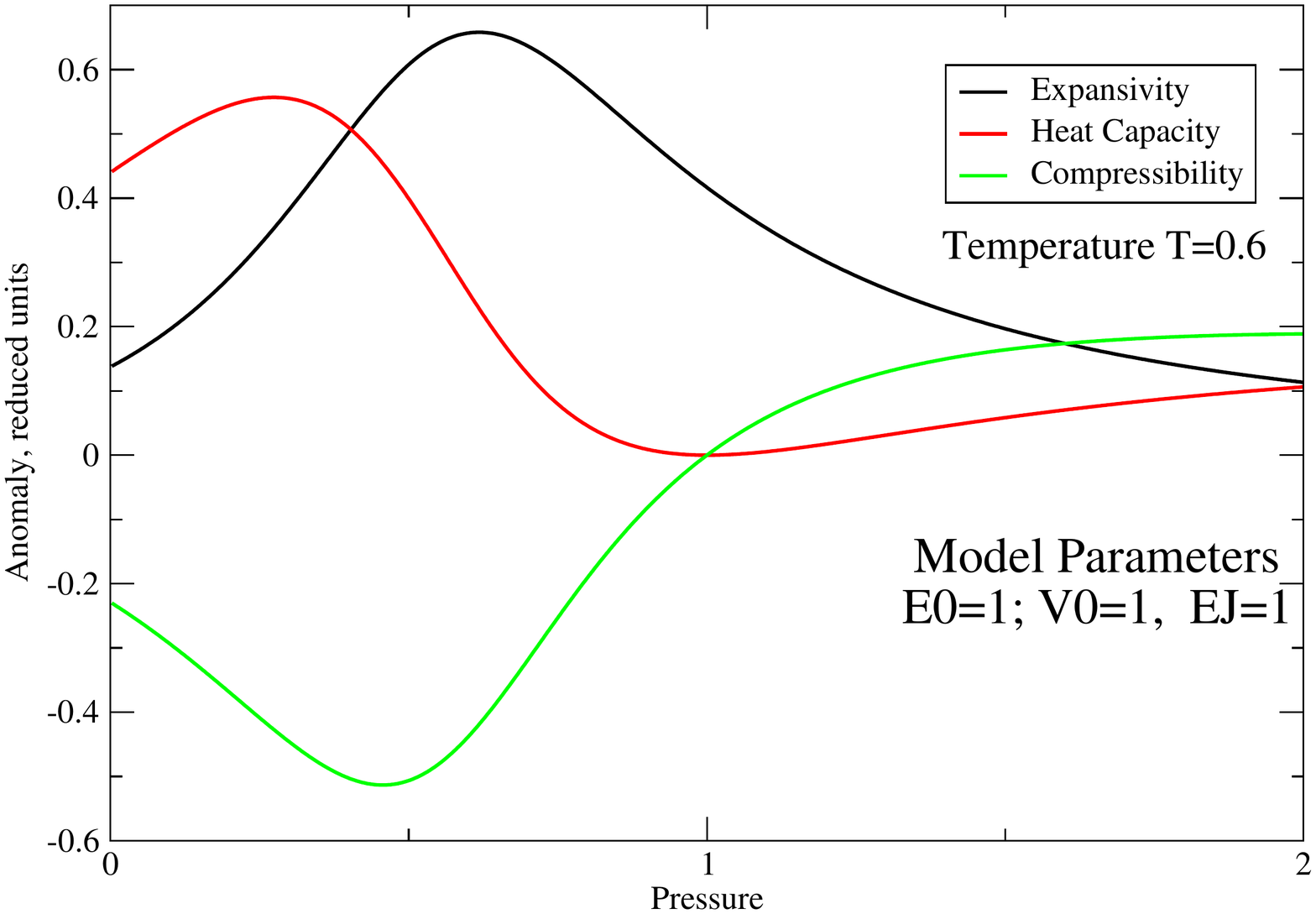}
\caption{Thermodynamic anomalies due to Bragg-Williams-type solid model (top) and liquid model (bottom) for T=0.4 and T=0.6.  with parameters set to unity, so that the solid critical temperature is 0.5 and the phase line is vertical at P=1  }
\label{fig:BW}
\end{figure}

These quantities are plotted in Figure~\ref{fig:BW}, where it is again clear that the model predicts a peak in specific heat and compressibility, along with a dip in the thermal expansivity. These extrema trace out the Widom lines of the phase diagram (Figure~\ref{fig:Widom}).  It is important to note that this phase diagram includes only the two-site Hamiltonian: the underlying free energy of the $x=1$ and $x=0$ states is ignored.

Free energy variation with $x$ is  shown in Figure~\ref{fig:BWx}, for a range of pressures around the phase transition. Below $T_c$ there are two minima, degenerate at  $P=\Delta U_e/ \Delta V_e$, indicating a first order phase transition. An analytic estimate of $T_c$ can be obtained from $\frac{\partial^2 G}{\partial x^2}=0$ or from setting $x=1/2/$.  Note that the existence of the critical point requires two nonlinear terms in $G$, coming here from the entropy and the interaction energy.   Figure \ref{fig:BWx}(lower) showing that $x$ changes discontinuously along an isotherm at the transformation, in either solid or liquid model. Notice that, for equivalent parameters, the critical point in the liquid falls at a lower $T_c$ than for the solid.

 If the liquid structure cannot accommodate enough potential electride sites, the model can be extended to a maximum electride fraction $f$. This would result in a change of the $P\Delta V_e$ prefactor from $x(2-x)$ to $x(2-x/f)$, but this additional complication makes no difference to the general argument, so hereinafter we take $f=1$.

Positive $J$ generates a first order transition with a  critical point.  The phase line is vertical (at $P=\Delta U_e/ \Delta V_e$) and ends at the critical temperature $Tc$.  Note that the high pressure phase transition we are describing corresponds to the Ising spin-up $\rightarrow$ spin-down transition, not the usual BW paramagnetic one.

Above the critical temperature, there are anomalies in several observables, as shown in Figure~\ref{fig:BW}. The extreme values  (Widom lines) for various properties do not fall in the same place: any definition of the supercritical transition pressure depends on which property is considered. 

\clearpage

\subsection{Entropy-driven transformation}
So far we have considered models where the difference between the two phases is in the enthalpy. In other cases, such as the molecular-atomic transition in hydrogen, there is a significant change in entropy between the two states - in the hydrogen case because the number of particles doubles.

The addition of an $x$-dependent entropy term gives a slope to the phase boundary, and a similar change of slope to the Widom lines: some of which can even have the opposite slope to the phase boundary.
From the Clausius-Clapeyron equation, the slope of the phase boundary is $\frac{dP}{dT} = \Delta S/\Delta V$.   Exactly similar to the volume change, a linear dependence of entropy with $x$ does not create  a first order transition, the lowest order term which can do so is $x(1-x)\Delta S$.  Such entropic demixing occurs in models with hard-core cubes and spheres  \cite{dijkstra1994phase,dijkstra1994evidence,dijkstra1998phase}, and has been claimed experimentally in supercooled water \cite{holten2012entropy}.

\begin{figure}[ht]
\centering
\includegraphics[width=0.8\columnwidth]{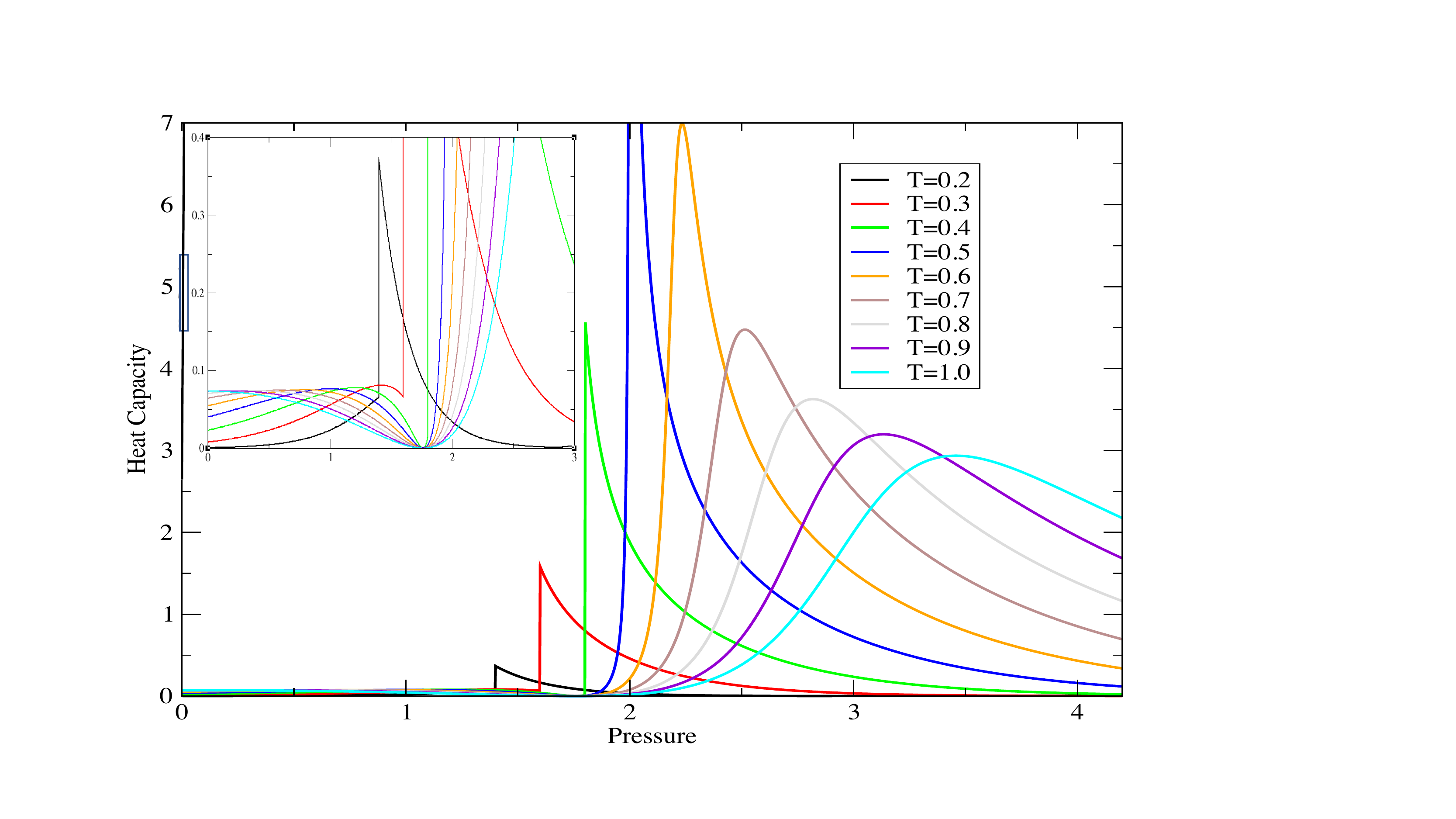}
\caption{Thermodynamic anomalies in the heat capacity for solid model with $\Delta V=1$, $\Delta U=1$, $J=1$, $\Delta S=2$.  Inset shows expanded view of low pressure region.  }
\label{fig:heatcapacity}
\end{figure}

The heat capacity model with $\Delta S=2$ is illustrated in Figure~\ref{fig:heatcapacity}, showing the lambda profile of the discontinuous transition changing to the broad peak above the critical point.  The gradient of the Clapeyron slope is evident from the shift of the lambda peak to higher pressure.  The inset shows the peak in heat capacity in the low-$x$ phase.

\subsection{Two site model for liquid-solid transformation}

We can extend the two-site model to compare liquid and solid phases and calculate a melt line. This requires us to consider  the $x$-independent contributions to the free energy.  A full equation of state is required for the non-anomalous contributions to $C_p$, $\alpha$ and $\kappa$, however, to calculate the phase boundary, we need only know the free energy {\it difference} of  $x$-independent contributions to the solid-liquid free energy $\Delta G_{sl}$.

Thus we have an equation for the phase boundary

\begin{equation} \Delta G_{sl}(P,T)=G^l(x_l,P,T)-G^s(x_s,P,T)\end{equation}

where $x_l(P,T)$ and $x_s(P,T)$ are the equilibrium values of $x$ in liquid and solid respectively

In figure \ref{fig:melt} we show an illustrative example
with a zero pressure melting point at T=0.4 and a positive Clapeyron slope of 0.8, in reduced units.  

To illustrate the model, we use the same parameters  $x$-dependent parameters $\Delta V=1$, $\Delta U=1$, $\Delta S=0$, $J=1$ in both liquid and solid. 
This means that  $x$-dependent terms in free energy for solid and liquid models are equal in the $x\rightarrow 0$ and $x\rightarrow 1$ limits.
 For the $x$-independent terms, we assume that energy, entropy and density differences between solid and liquid are constant.

This choice of parameters means that the solid-solid phase line is vertical ($\Delta S=0$) and there is no discontinuous liquid-liquid transition.  This is similar to the case of the simple metals.   A significantly larger value of $J$ would be needed to  extend the phase boundary into the liquid region, as shown in figure 3.
A non-zero $\Delta S$ leads to a slope in the phase boundary, but does not change the general picture.

The figure also shows how $x$ varies across the phase diagram - gradually in the liquid, but discontinuously in the solid.

\subsection{Example - application to potassium}

The high pressure crystallography and re-entrant melt curve of potassium has been determined experimentally\cite{Lundegaard2009b,mcbride2015one,jr:Reentrantincommensuratepotassium}. DFT calculations show an electride transition in potassium as in other simple metals\cite{falconi2006ab,marques2009potassium,gatti2010sodium,woolman2018structural,Marques2011,marques2011optical,zhao2019commensurate,Miao2014}.  Liquid potassium calculations suggest a number of irregularities\cite{zongK} in the  thermodynamic properties which cannot be fitted by smoothly varying models\cite{li2019regularities}.   

\begin{figure}[ht]
\centering
\includegraphics[width=0.92\columnwidth]{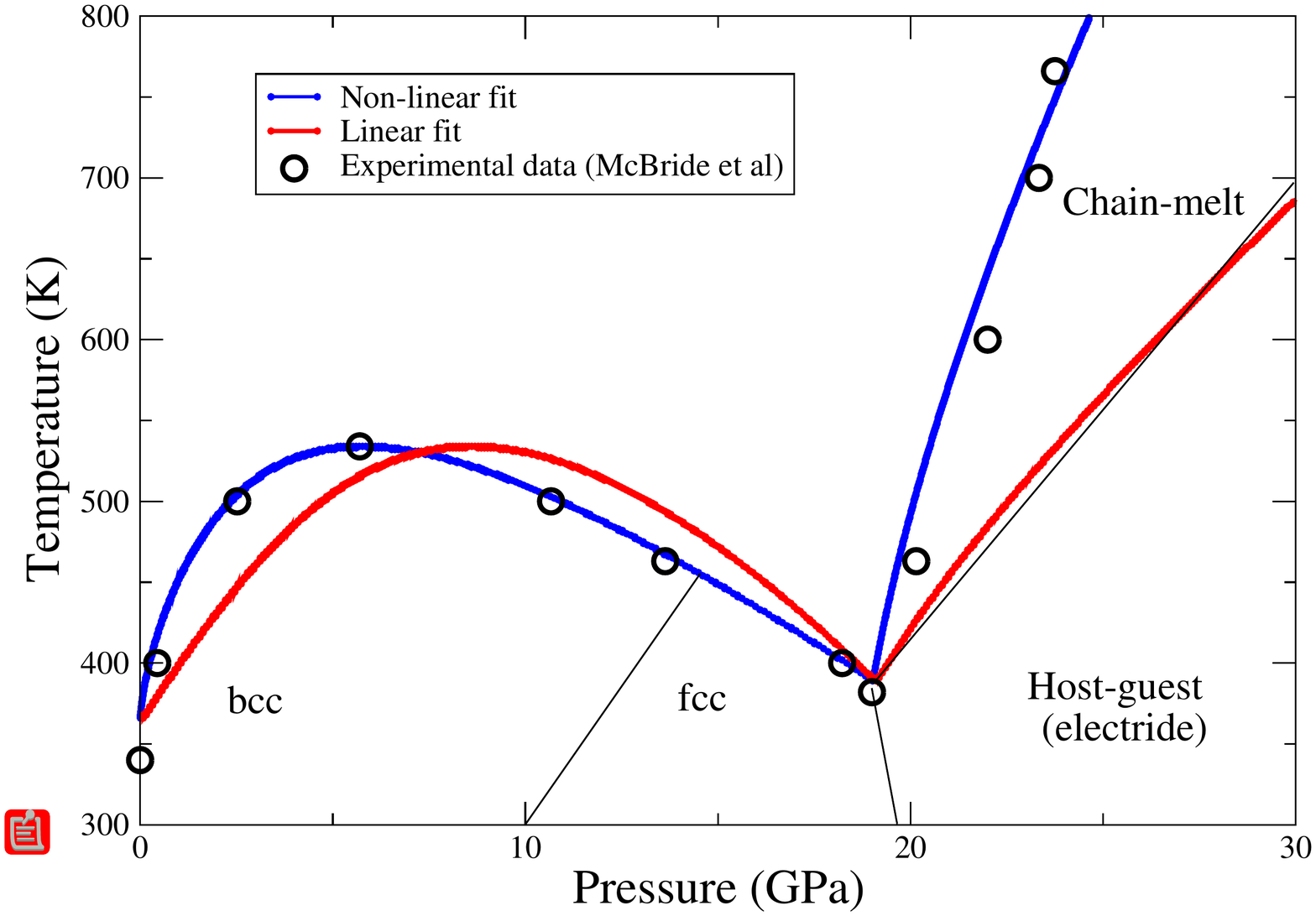}
\includegraphics[width=0.98\columnwidth]{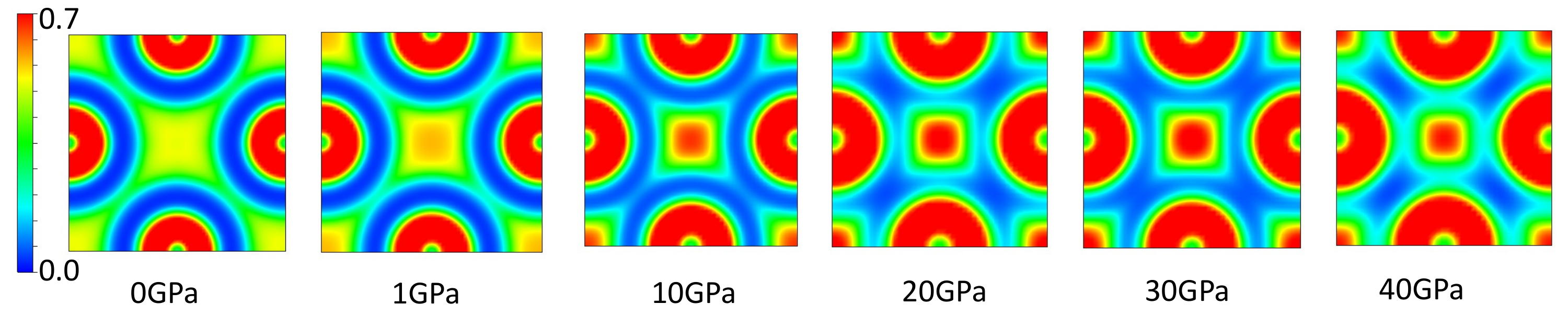}
\includegraphics[width=0.98\columnwidth]{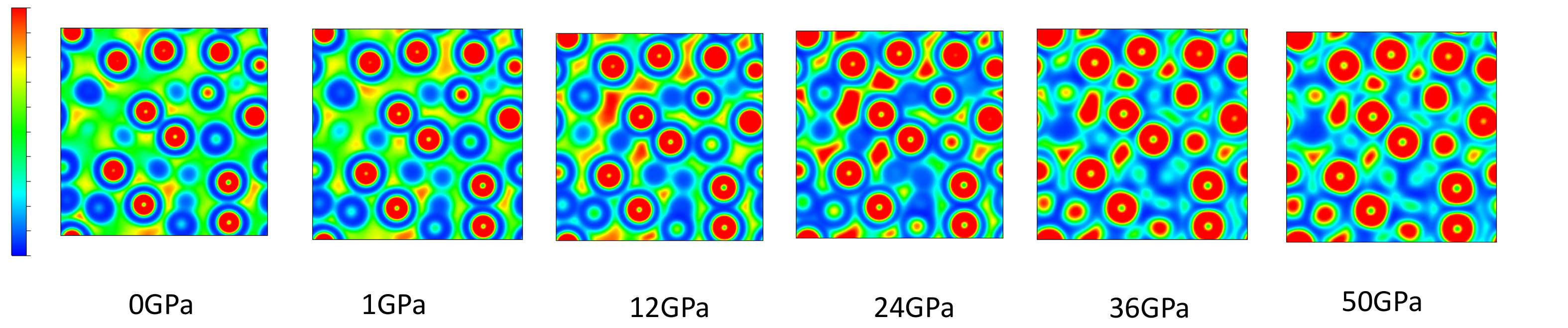}
\caption{Model fitted to potassium. (upper) The phase diagram based on two state model used parameters $\Delta U=1.15$, $\Delta V=1$, $J=1$ , $\Delta S=-0.5$, the same in both solid and liquid phases.  The $x$-independent free energy difference between solid and liquid phases are
$\Delta G_{sl}= 0.12+0.24P-0.33T$.
(lower) Sequence of images showing the transition to electride state in fcc and liquid potassium.  Figure shows electron localisation function\cite{schmider2000chemical} (ELF) calculated using CASTEP density functional code\cite{jr:CASTEP}. 
Red shows region of high ELF and localised charge (see scale bar for values).  The solid electride site is located at the ($\frac{1}{2},\frac{1}{2},\frac{1}{2})$ position - centre of the figure.  The "liquid" image is a slice from a molecular dynamics snapshot\cite{zongK} cutting through several atomic sites (red circles at 0GPa - not all lie in the plane.), and then rerun at several densities using the same fractional atomic positions. the intersitial regions, initially green (free electronlike) become increasing red and blue as the electrons localise. 
\label{fig:KLF}}
\end{figure}

\begin{figure}[ht]
\includegraphics[width=0.88\columnwidth]{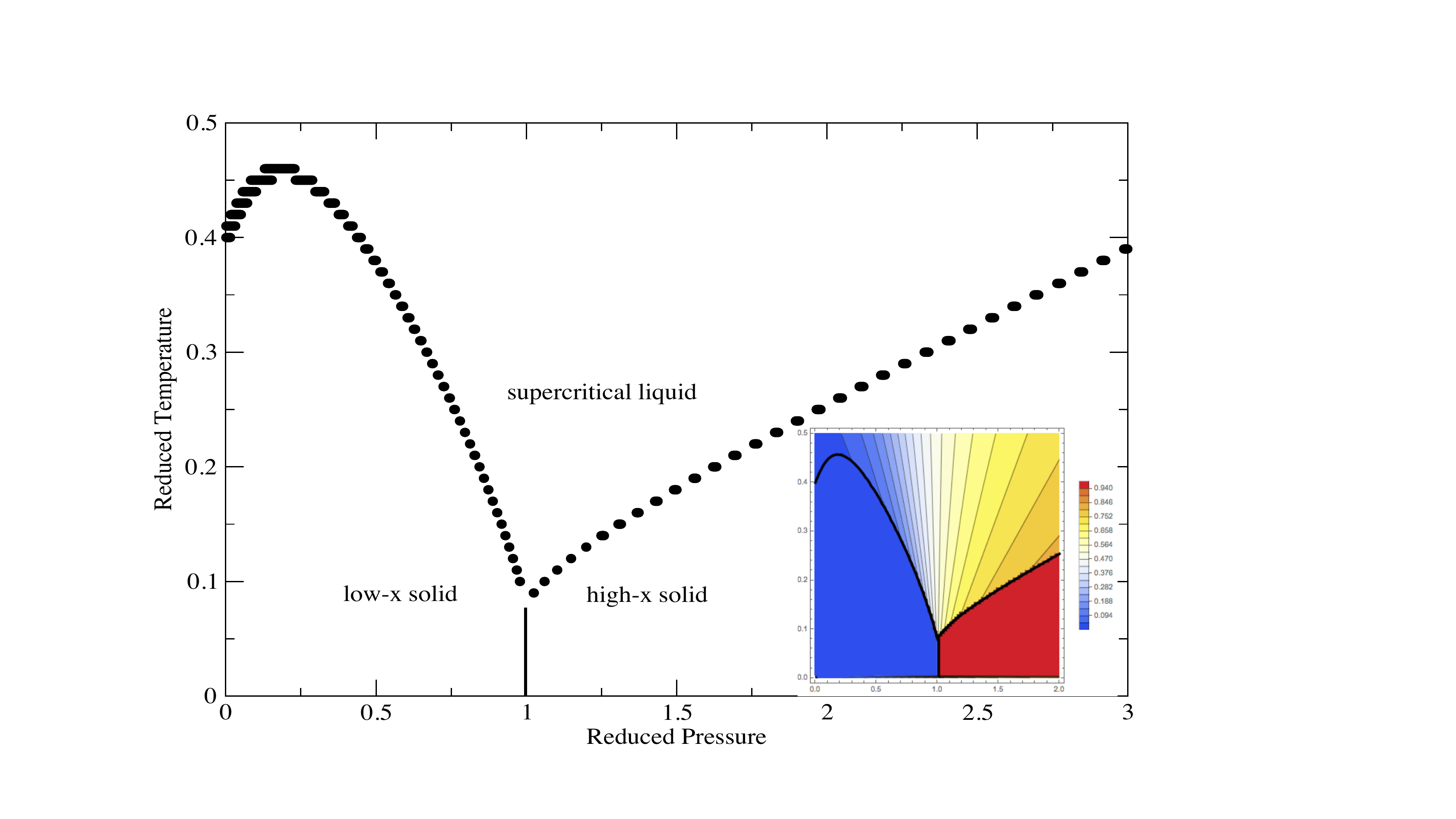}
\caption{Phase diagram for the combined solid-liquid two-site model with $\Delta V=1$, $\Delta U =1$, $\Delta S =0$, J=1 in both phases, and a linear free energy difference $\Delta G_{sl}=0.02+0.04P-0.05T$
for the non-anomalous contributions.  Data was collected by scanning a dense grid in P-T space and plotting points where the free energy difference was less than 0.0005, in reduced units. Inset shows variation of x across the phase diagram, from blue (x=0) to red (x=1) }
\label{fig:melt}
\end{figure}

Figure \ref{fig:KLF} shows an {\it ab oculo} parameterization of the liquid-solid transformation model to this data, with a simple linear model for $x$-independent terms.  The reduced units of the model, correspond to 20GPa and 1000K for potassium.
While the overall shape is reproduced with a linear fit, the low pressure melt curve appears parabolic and the high pressure line is not sufficiently steep.  The fit can be significantly improved by introducing a non-linear equation of state, such that $P^*\rightarrow P^{3/2}$, and is reduced by a factor of 5 above the transition.

Curiously, the unadjusted high-pressure melt line of the linear model follows the chain-melting line, in which the guest atoms in the solid phase III melt\cite{chainmeltingpotassium,mcbride2015one}.

It is notable that the melting line minimum is coincident with the triple point of the solid-solid transformation from fcc to host-guest structure, which has been associated with the electride transition\cite{marques2009potassium,woolman2018structural,robinson2019chain}.  The melting point maximum has no such association, which casts doubt on the extrapolation of the fcc-bcc line to the melt curve maximum, which has been drawn and copied without evidence in, e.g. lithium\cite{guillaume2011cold,ackland2017quantum,Marques2011,jr:Lithium,Hanfland2000,schaeffer2012high}.  In fact, the 180$\deg$ rule means that  it is thermodynamically impossible for a solid-solid phase line to intercept the melt curve at a point of negative curvature such as a maximum.

\section{Discussion}

We have presented a simple analytic model which explains the anomalous shape of the melt line observed in many high pressure systems.  The key features required are 
\begin{itemize}
\item a microscopic mechanism by which the atoms can reduce their volume, at the expense of increasing their energy.   \item disorder in the liquid leading to wider range of possible atomic environments compared to the solid.  
\end{itemize}

We have shown that a discontinuous phase transformation can be driven by a repulsion between the two states, analogous to the $Jx(1-x)$ term in the Bragg-Williams model. This repulsion may be either enthalpic or entropic, but much introduce negative curvature to $G(x)$: terms linear in $x$ cannot result in a discontinuous transition.

The discontinuous transitions in the model do not depend on changes in crystal symmetry; In reality, it is likely that a discontinuous change in the type of electronic binding of a solid will also be accompanied by a symmetry change.  Thus even in principle the solid-solid critical point can occur only for isostructural transitions such as hydrogen and cerium\cite{ji2019ultrahigh,ackland2020structures,johansson1974alpha}.

Above the critical point, the model predicts a series of experimentally measurable "Widom lines" associated with anomalies of thermodynamic properties.  These occur for all parameterizations, even where there is no critical point, or there is a liquid-liquid critical point which lies below the melt line.  

By comparing free energy models for crystal and liquid phases, we constructed a melt line from this model.   This has a characteristic minimum at the point where the two-state mixing entropy is maximised ($x=\frac{1}{2}$), coincident with the solid-solid phase transformation.  Combined with a positive slope at low pressures, this means that there must also be a melting temperature maximum which, curiously, does not appear to be coincident with other features in the phase diagram.

The model has been applied to the melt curve of potassium, using a very simple linear fitting scheme. More accurate fitting to other materials would be straightforward, and the model framework has broad application for producing equations of state for any material with a complex liquid-liquid transformation.

  \acknowledgements{HZ and GJA acknowledge the ERC project HECATE for funding. HZ thanks the National Natural Science Foundation of China (51931004 and 51871177).  Computing resources were obtained via the UKCP EPSRC grant EP/P022561/1.  We thank X. Ding and H Ehteshami for useful discussions.}

\clearpage

\bibliographystyle{apsrev}
\bibliography{references}

\clearpage
\appendix{Python code to solve the two site liquid system}

\end{document}